\newcommand{\CLASSINPUTtoptextmargin}{1.85cm}
\newcommand{\CLASSINPUTbottomtextmargin}{2.49cm}
\documentclass[conference]{IEEEtran}
\IEEEoverridecommandlockouts
\usepackage{cite}
\usepackage{amsmath,amssymb,amsfonts}
\usepackage{bm}
\usepackage{acronym}
\usepackage{algpseudocode}
\usepackage{algorithm}
\usepackage{graphicx}
\usepackage{textcomp}
\usepackage{xcolor}
\usepackage{subfig}
\usepackage{pgfplots}
\usepgfplotslibrary{groupplots,dateplot}
\usetikzlibrary{patterns,shapes.arrows}
\pgfplotsset{compat=newest}
\usepackage{dsfont}
\def\BibTeX{{\rm B\kern-.05em{\sc i\kern-.025em b}\kern-.08em
    T\kern-.1667em\lower.7ex\hbox{E}\kern-.125emX}}
\algnewcommand\algorithmicforeach{\textbf{for each}}
\algdef{S}[FOR]{ForEach}[1]{\algorithmicforeach\ #1\ \algorithmicdo}

\newcommand\copyrighttext{%
  \footnotesize \textcopyright 2024 IEEE. Personal use of this material is permitted. Permission from IEEE must be obtained for all other uses, in any current or future media, including reprinting/republishing this material for advertising or promotional purposes, creating new collective works, for resale or redistribution to servers or lists, or reuse of any copyrighted component of this work in other works.}
\newcommand\copyrightnotice{%
\begin{tikzpicture}[remember picture,overlay]
\node[anchor=south,yshift=10pt] at (current page.south) {\fbox{\parbox{\dimexpr\textwidth-\fboxsep-\fboxrule\relax}{\copyrighttext}}};
\end{tikzpicture}%
}

\begin{document}

\title{Decentralized LLM Inference over Edge Networks with Energy Harvesting

\thanks{This work has been supported by the EU H2020 MSCA ITN project Greenedge (grant no. 953775), by the EU through the Horizon Europe/JU SNS project ROBUST-6G (grant no. 101139068), and by the EU under the Italian National Recovery and Resilience Plan (NRRP) of NextGenerationEU, partnership on “Telecommunications of the Future” (PE0000001 - program “RESTART”).}
}

\author{\IEEEauthorblockN{Aria Khoshsirat, Giovanni Perin, and Michele Rossi}
\IEEEauthorblockA{\textit{Department of Information Engineering (DEI)} \\
\textit{University of Padova (Padova, Italy)}\\
Emails: aria.khoshsirat@unipd.it, giovanni.perin.1@unipd.it, michele.rossi@unipd.it}
}

\maketitle
\copyrightnotice

\acrodef{LLM}{large language model}

\begin{abstract}

Large language models have significantly transformed multiple fields with their exceptional performance in natural language tasks, but their deployment in resource-constrained environments like edge networks presents an ongoing challenge. Decentralized techniques for inference have emerged, distributing the model blocks among multiple devices to improve flexibility and cost effectiveness. However, energy limitations remain a significant concern for edge devices. We propose a sustainable model for collaborative inference on interconnected, battery-powered edge devices with energy harvesting. A semi-Markov model is developed to describe the states of the devices, considering processing parameters and average green energy arrivals. This informs the design of scheduling algorithms that aim to minimize device downtimes and maximize network throughput. Through empirical evaluations and simulated runs, we validate the effectiveness of our approach, paving the way for energy-efficient decentralized inference over edge networks.

\end{abstract}

\section{Introduction}
The recent and rapid growth of large language models (LLMs) has transformed various fields by achieving remarkable performance in natural language understanding and generation tasks~\cite{llmsurvey}. From language translation~\cite{llmtranslation} to content recommendation systems~\cite{llmrecommender}, LLMs have become indispensable tools to tackle complex linguistic challenges. However, as the capabilities and complexities of these models keep increasing, so do the demands for computational resources, posing significant challenges for their deployment in real-world applications.

One of the key challenges in the implementation of LLMs lies in the efficient utilization of computational resources~\cite{llmresource}, particularly in resource-constrained environments such as edge networks. 
Edge networks, which consist of devices located close to data sources, provide benefits such as decreased latency, increased privacy, and improved reliability~\cite{edgesurvey}. However, the constrained energy and memory capacity of edge devices pose significant challenges to the implementation of resource-intensive models such as LLMs. To address these challenges, decentralized techniques for LLM inference are gaining traction~\cite{llmdecentralized1,llmdecentralized2,llmdecentralized3}. These techniques offer several advantages over traditional centralized approaches by distributing the workload across multiple devices. Specifically, they provide greater flexibility and cost-effectiveness while alleviating the burden on individual devices by harnessing the computational power of multiple devices in a decentralized manner.

Despite the advantages offered by decentralized LLM inference, the energy constraints of edge devices remain a critical concern. The limited power available to edge devices requires innovative solutions to ensure sustainable operation, particularly in scenarios where reliable and continuous inference is needed. Energy harvesting techniques, which capture and convert ambient energy sources such as solar, kinetic, or thermal energy into electrical energy, show potential to supply energy for modern edge computing tasks~\cite{wei2018enabling, energyharvesting2}. By integrating energy harvesting capabilities into edge devices, we can increase their energy reserves and extend their operational lifetime, thereby facilitating the deployment of LLMs in energy-constrained environments. In addition, the use of renewable energy sources aligns with the broader objectives of environmentally friendly computing practices, contributing to the reduction of the carbon footprint associated with computational tasks.

This study explores the decentralized LLM inference over edge computing networks. Our research aims to investigate the integration of energy harvesting devices with edge computers for sustainable and efficient LLM inference. The contributions of this paper to this emerging field of study are as follows: 
\begin{itemize}
    \item Empirical energy measurements of LLM transformer blocks are performed on a commercial edge device (Nvidia Jetson AGX Orin) while implementing a network simulation for collaborative inference.
    \item A semi-Markov model describing the battery state of the edge computer and future evolution based on processing parameters and green energy arrivals is developed.
    \item Scheduling approaches for the distributed inference task are designed using the developed model to minimize device downtimes and maximize job throughput.
    \item A new dynamic power mode for the operation of edge devices is introduced to optimize the balance between energy consumption and task completion time. 
\end{itemize}

These efforts collectively contribute to the development of energy-efficient and scalable solutions for decentralized LLM inference over edge networks.

\section{Problem statement and Characteristics}
\label{sec:problem_statement}

\begin{figure}[t]
\centerline{\includegraphics[width=0.52\textwidth,keepaspectratio]{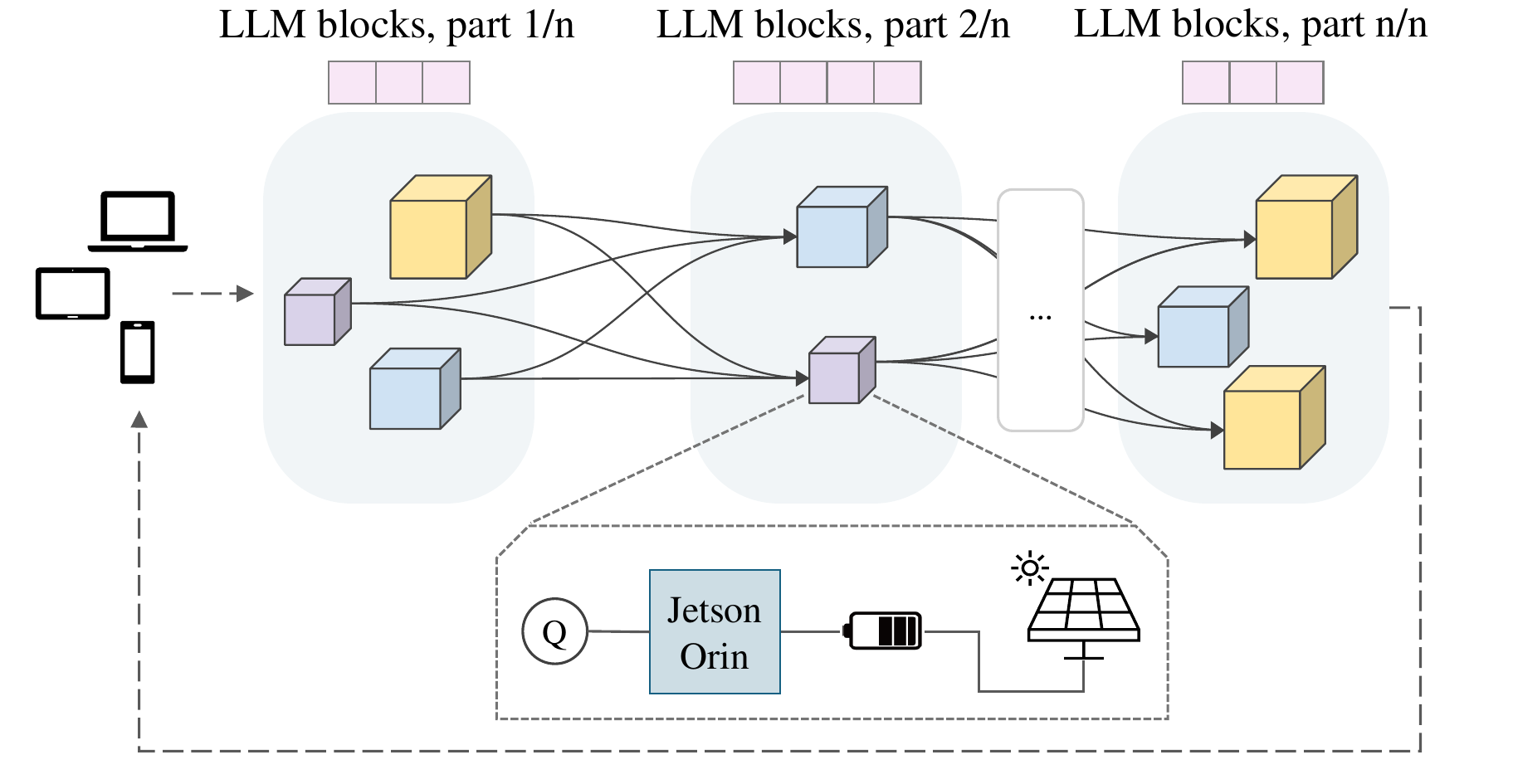}}
\centering
\caption{\textbf{Decentralized inference network architecture.} Each cube represents a node comprising a battery-powered edge device linked to an energy source.}
\vspace{-0.3cm}
\label{fig:setup}
\end{figure}

Building upon \textit{Petals}~\cite{llmdecentralized2}, a framework for decentralized LLM inference across interconnected devices, we introduce novel considerations to accommodate battery-powered edge devices. Like \textit{Petals}, our framework distributes the LLM layers between groups of devices, and, within a group, identical portions of the LLM layers are replicated on each device, enabling parallel inference. Fig.~\ref{fig:setup} shows an overview of the framework. When a new request from an end user arrives (referred to as a \textit{job} in subsequent sections), a device from each group is designated to perform an LLM inference run. The chosen device within each successive group receives the output from the preceding group as its input and transmits the output generated by its LLM blocks to a device in the subsequent group.

We consider each device within the decentralized inference framework as a battery-powered edge device. The example edge device utilized in our experiments is the Jetson AGX Orin~\cite{orin}, a state-of-the-art commercial edge device from Nvidia. Orin, the latest addition to the Jetson series, stands out as the most advanced platform, featuring 12 CPU cores, a powerful GPU, and ample memory capacity. It can operate in four distinct power modes, specifically at 15, 30, 50, and 60 watts. These hardware specifications make it well suited to execute resource-intensive tasks on the edge~\cite{orin2} with high energy efficiency. These devices can be connected to renewable energy sources, such as solar panels, to mitigate energy constraints and enable sustained operation. The energy generated by solar panels can fluctuate over time, although it tends to remain relatively stable in short intervals. Therefore, we characterize the energy arrival rates as samples from a uniform distribution bounded by two values specific to each node, for every time step. This approach acknowledges the unique energy profiles of individual edge devices within the network, which may stem from variations in solar panel models or weather conditions at their respective locations.

Each edge device can accommodate a single job in its queue $Q$, representing a pending task awaiting processing. This limitation to one job in the queue is particularly relevant for inference in LLMs and caching mechanisms, which often require significant memory resources. Considering the limited memory capacity of edge devices, restricting the queue to one job helps prevent memory overflow and promotes optimal resource utilization. Devices can enter a \emph{downtime period} where they are put in \emph{power-saving mode}. This occurs when a low battery threshold is hit until an upper threshold of safe battery level is recovered. This ensures that the devices do not deplete their battery entirely, shutting down as a result. This is particularly relevant considering that edge devices are typically employed for various tasks and should ideally remain operational without running out of energy.

Our primary objective is to optimize resource utilization across a heterogeneous edge network setup, considering variations in device capabilities such as energy consumption, processing time, and the rate of energy replenishment. This entails intelligently and efficiently selecting devices within each group for every inference run, aiming to minimize the downtime of these devices and maximize the overall network computing throughput. Furthermore, we investigate switching between different power modes implemented in edge devices, focusing on those applicable to the Jetson AGX Orin, considering changes in energy consumption and inference time. Through this comprehensive examination, our aim is to provide insight into the optimal utilization of power modes to improve both energy efficiency and the task completion rate in edge computing environments.

\section{System Model for Edge Devices}
\label{sec:single_node_modeling}

As mentioned in the previous section, we consider each edge device to be connected to a renewable energy source, e.g., a small form factor solar panel. In addition, a local battery is also included to store the collected energy. The system operates according to discrete time slots with fixed duration $\delta$ (the reference time unit). New job arrivals obey a Bernoulli model, where a new job arrives in any time slot according to a fixed probability $p$. For the energy model, we assume that the renewable source generates $e \geq 0$ units of energy in a time slot according to a discrete mass distribution function (MDF) $f(e)$. Energy arrivals in different time slots are assumed to be statistically independent.

The operation of any edge node $j$ is described through consecutive {\it stages} defined by the variable \mbox{$m=0,1,2,\dots$}. In each processing stage $m$ there are several options: {\it i)} if the device is in the active state and the processing queue is non-empty, a new job is processed, {\it ii)} if the device is active but no jobs are available, it will be idling, and {\it iii)} if it is in the so-called power saving mode, the computation is momentarily suspended until the energy level in the battery will increase over a certain threshold. Stages have a duration that is a multiple of $\delta$, namely, $\kappa \delta$, where $\kappa \geq 1$ is an integer depending on the power mode, through the (binary) power saving mode variable $PM$. The power saving state of the node corresponds to $PM=0$, and the active states have $PM\geq 1$. The duration of stage $m$ corresponds to the time taken to process the job at that stage, which directly descends from the chosen power mode. The number of slots in stage $m$ is referred to as $\kappa_m$. Due to statistical independence, the MDF of the energy inflow in stage $m$ is obtained by convolving $k_m$ times the MDF $f(e)$. 

At edge node $j$, the energy $E^j_m$ that is available in its battery at the beginning of stage $m$ is expressed as a multiple of a reference energy unit, with $E^j_m = 0,1,\dots, E_{\max}$, where $E_{\max}$ represents the maximum number of energy units that can be stored in the battery. At the beginning of stage $m+1$, the battery level $E^j_{m+1}$ is updated as

{\small
\begin{equation}
\label{eq:def_energy_per_sample}
    E^j_{m+1} = \max\left(\min\left(E^j_{m} + \Delta IE_{m} - CE(PM),E_{\max}\right), 0\right)
\end{equation}
\par}

where $E^j_{m}$ is the energy in the battery at the beginning of stage $m$, $\Delta IE_{m}$ is the energy collected from the local energy source during stage $m$, and $CE(PM) \leq E^j_{m}$ denotes the energy consumed by node $j$ for computation in stage $m$.

Each edge node $j$ stores the jobs that are to be processed in a local queue, and this queue at stage $m$ is denoted by $Q_m^j$.
In what follows, we refer to the generic edge node $j$ omitting the index for the sake of clarity. The state of the edge node in stage $m$ is described by the tuple $S_m=(Q_m, E_m, \gamma_m)$. This state contains the queue state $Q_m~\in~\{0,1\}$, the discrete energy level $E_m$ and a binary variable $\gamma_m \in \{0,1\}$ that indicates whether the node is in ``power saving'' or ``active'' mode. Together, these variables form the state space of a semi-Markov chain~\cite{semimarkov} describing the edge node evolution and play a key role in describing the system behavior and transitions at any specific time step $m$. This state representation enables us to analyze the statistical evolution of the edge node over time. %

At any stage $m$, a power mode ($PM$) is set for each edge node, in the power saving mode, it holds $\gamma_m=PM=0$ and in any state with $\gamma_m=0$ new jobs are rejected. The node is put into power saving if $E_m$ becomes smaller than a (user-defined) threshold $E_{\rm th}$ and it remains in power saving state until the energy $E_m$ increases beyond a second threshold $E^\prime_{\rm th} > E_{\rm th}$. When this occurs, $\gamma_m$ is set back to $\gamma_m=1$ to indicate that the node is back to ``active''. This enforces a hysteresis on the battery level to prevent unwanted oscillations in the node power saving behavior. In all active states ($\gamma_m=1$) the power mode $PM \geq 1$ is deterministically set (see Section~\ref{sec:results}) depending on the energy level $E_m$: this reflects the fact that different configurations (processing power and speed) can be used depending on the available energy. In the active states ($\gamma_m=1$), new job arrivals are accepted as follows: we consider that a new job arrives during stage $m$ if a new arrival occurs in at least one of the time slots within this stage. This event probability is $p_m = 1-(1-p)^{\kappa_m}$, where the stage duration $\kappa_m$ is deterministically obtained from $PM$. %

The transition probabilities between the edge node states, \mbox{ $S_m=(Q_m, E_m, \gamma_m) \rightarrow S_{m+1}=(Q_{m+1}, E_{m+1}, \gamma_{m+1})$} are computed as follows:

\begin{itemize}
\item If $Q_m = Q_{m+1} = 0$, there are no jobs at the node, neither pending nor being computed. Therefore, all the feasible transitions must have $E_{m+1} \ge E_{m}$. Moreover, we discriminate between two cases:
\begin{itemize}
    \item[1)] If $\gamma_m=1$, the allowed transitions are to state $S_{m+1} = (0, E_{m+1}, 1)$, with $E_{m+1} \geq E_m$ and transition probability $P(\Delta IE_m = E_{m+1}-E_{m}) \times (1-p_m)$, where $P(\cdot)$ indicates a probability. 
    \item[2)] If $\gamma_m=0$ the allowed transitions are to state $S_{m+1} = (0, E_{m+1}, \gamma_{m+1})$, where $E_{m+1}\geq E_m$, with probability $P(\Delta IE_m = E_{m+1}-E_{m})$, where $\gamma_{m+1}=0$ if $E_{m+1} \leq E^\prime_{\rm th}$ and $\gamma_{m+1}=1$ otherwise (hysteresis). The arrival probability in this case is irrelevant as new arrivals are rejected when $\gamma_m=0$.
\end{itemize} 
\item If $\gamma_m=0$ and $Q_m=1$ the pending job is not computed and new job arrivals are ignored. The transition to the next state $(1, E_{m+1}, \gamma_{m+1})$ has probability $P(\Delta IE_m = E_{m+1}-E_{m})$. Moreover, $\gamma_{m+1}=0$ if $E_{m+1} \leq E^\prime_{\rm th}$ and $\gamma_{m+1}=1$ otherwise.

\item If $\gamma_m=1$, $Q_m=0$ and $Q_{m+1} = 1$, the job queue is empty and a new job arrives in stage $m$. Again, all the feasible transitions towards stage $m+1$ have $E_{m+1} \ge E_{m}$, since there is no job being processed in stage $m$. The transition probability to state $S_{m+1} = (1, E_{m+1}, 1)$ is \mbox{$P(\Delta IE_m=E^j_{m+1}-E^j_{m}) \times p_m $}.

\item If $\gamma_m=1$, $Q_m=1$ and $Q_{m+1} = 0$, the node in stage $m$ will be processing the pending job in the queue and no new jobs arrive at the node during stage $m$. The transition probability towards state $S_{m+1}$ in this case is \mbox{$P(\Delta IE_m=E_{m+1}-E_{m}+CE(PM)) \times (1-p_m)$}. The node will transition to state $S_{m+1} = (0, E_{m+1}, \gamma_{m+1})$, where $\gamma_{m+1}=0$ if $E_{m+1} < E_{\rm th}$ and $\gamma_{m+1}=1$ otherwise.

\item If $\gamma_m=1$, $Q_m = Q_{m+1} = 1$, the node will be processing an accepted job in stage $m$, and in the same stage a new job will also arrive. The transition probability to state $S_{m+1} = (Q_{m+1}, E_{m+1}, \gamma_{m+1})$ in this case is \mbox{$P(\Delta IE_m=E_{m+1}-E_{m}+CE(PM)) \times p_m$}, where $\gamma_{m+1}=0$ if $E_{m+1} < E_{\rm th}$ and $\gamma_{m+1}=1$ otherwise.
\end{itemize}

Let $T_S = \kappa \delta$ be the dwell time of a state $S$. The dwell time of those states where $Q_m=0$ corresponds to a single time slot ($\kappa=1$). The dwell time of those states where $\gamma_m=1$ (i.e., $PM \geq 1$) and $Q_m=1$ corresponds to the computational time $\kappa_m$ for a single job, which depends on the power mode $PM$ of that state. The dwell time of a state for which $PM=\gamma_m=0$ (power saving) also corresponds to a single time slot: in this case, new jobs are rejected, and the node keeps idling until the energy increases above $E^\prime_{\rm th}$.

Given this semi-Markov model, several key performance measures can be computed. For instance, the average energy level of a node in a time slot is calculated as
\begin{equation}
\bar{E} = \frac{\sum_{S} \pi_S E_S}{\sum_{S} \pi_S T_S},
\end{equation}
where the summations extend over all states, $E_s$ represents the energy level in state $s$, and $\pi_s$ is the stationary distribution of the embedded Markov chain.  Another metric of interest is the total probability that, in any time slot, the energy of a device will be less than a chosen threshold $E_{\rm \textit{lim}}$, computed as
\begin{equation}
\label{eq:xi_lim}
\xi_{\rm \textit{lim}} = \frac{\sum_{S \text{ s.t. } E \leq \rm E_{\textit{lim}}} \pi_S T_S}{\sum_{S} \pi_S T_S}.
\end{equation}
By establishing a suitable threshold, such as the battery level that triggers the power-saving mode, this risk metric can be minimized, thereby minimizing device downtime. Note that the stationary distribution only needs to be computed once unless the network parameters change, making the algorithm cost-effective.

\section{Distributed Control of in-Network Processing}
\label{sec:distributed_control}

We devise three scheduling algorithms of increasing complexity, selecting the job assignment probability distribution in different ways. Specifically, the
%aria
proposed policies are:
\begin{itemize}
    \item\textbf{Uniform.} The target edge device is chosen by sampling a uniform random variable among the devices that are currently available for processing.
    \item\textbf{Long-term.} The scheduling probability distribution is selected to minimize the risk of the battery level falling below a given threshold, using the developed semi-Markov model. This policy yields a static optimal-on-average solution over an infinite horizon.
    \item\textbf{Adaptive.} The long-term-obtained processing input rates are further refined in an adaptive way considering the current (instantaneous) energy state. The idea is that a part of the input probability is moved from devices with low energy to devices with higher energy availability.
\end{itemize}
\vspace{0.1cm}

\noindent\textbf{Algorithm description.} To start with, we define the shorthand notation $PM_i$ if $PM=i$ and with $PM(d)$ we mean the power mode of device $d$. Algorithm~\ref{alg:split_inference} summarizes the system operation and the above scheduling policies. At the beginning of each time slot, the energy states of the devices are updated according to Eq.~\eqref{eq:def_energy_per_sample} (line \ref{line:energy_update}), and the device $d$ chooses its power mode $PM(d)$ according to a pre-defined lookup table (line \ref{line:lookup}, see Section~\ref{sec:results}). Then, the scheduling probabilities are set for each layer $\ell$ according to the adopted \texttt{scheduling} procedure (line~\ref{line:scheduling}), i.e., uniform, long-term or adaptive. After setting the scheduling probabilities, decentralized inference of the input can be performed. Since the stationary distribution is computed offline, scheduling subroutines can be executed fast, making the algorithm suitable for real-time applications.

\setlength{\textfloatsep}{4.5pt}

\begin{algorithm}[bt]
\caption{Split inference of LLMs}\label{alg:split_inference}
\small
\begin{algorithmic}[1]
\State\textbf{input:} device limit input rates vector $q_{\rm \textit{lim}}^\ell$ for every layer $\ell$.
\Repeat
\ForEach{device $d$}
    \State Energy update (Eq.~\eqref{eq:def_energy_per_sample}) \Comment Update the energy state
    \label{line:energy_update}
    \State $PM(d) \gets \texttt{choose}(PM_0,\ldots,PM_M)$ \Comment Select $PM$
    \label{line:lookup}
\EndFor
\ForEach{layer $\ell$}
    \State $q^\ell \gets$ \texttt{scheduling}($q_{\rm \textit{lim}}^\ell$)
    \Comment Set input probs.
    \label{line:scheduling}
\EndFor
\State \texttt{inference}(input) \Comment Run the system
\Until The system is working.
%aria
\Procedure{Uniform}{x}
\State $x \gets 1/\texttt{length}(x)$
\State \textbf{return} $x$
\EndProcedure
\Procedure{Long\_term}{x}
\State $x \gets x/\sum_i x_i$
\label{line:long_term_normalization}
\State \textbf{return} $x$
\EndProcedure
\Procedure{Adaptive}{x}
\State $x \gets$ \texttt{long\_term}($x$)
\label{line:long_term}
\ForEach{device $i^\ell$}
\If{device $i^\ell$ having power mode $PM_1$}
\State $z \gets \alpha / N_\ell$
\State $x_{i} \gets x_{i} - (1-z) x_{i}$ \Comment decrease input rate
\label{line:update_q}
\EndIf
\State $x \gets x/\sum_i x_i$ \Comment normalize to get a valid MDF
\label{line:normalization}
\EndFor
\EndProcedure
\end{algorithmic}
\end{algorithm}

\noindent\textbf{Static long-term optimal scheduling.} The long-term optimal job arrival probability is determined by analyzing the energy consumption of the edge devices under a known energy arrival distribution. Specifically, the maximum acceptable input rate $q_{\rm \textit{lim}}^{\rm energy}$ relative to the maximum tolerable risk $\xi_{\rm \textit{lim}}$ (user-defined) can be retrieved via a root-finding algorithm, such as Brent's method~\cite{brent}. The method is fed with the function given by Eq.~\eqref{eq:xi_lim}, where the stationary distribution $\pi_S$ depends on the variable $q$ (job input rate).
The obtained input rate $q_{\rm \textit{lim}}^{\rm energy}$ only considers the energy risk minimization, lacking the inclusion of time constraints due to processing. Let $\bar{\kappa}$ be the expected number of slots for processing a task, it holds
\begin{equation}
    \label{eq:T_S}
    \bar{\kappa} = \frac{\sum_{S \text{ s.t. } (Q=1, \gamma=1)} \pi_S \kappa_S}{\sum_{S \text{ s.t. } (Q=1, \gamma=1)} \pi_S}.
\end{equation}
When the power mode is fixed, the processing slots are also fixed, i.e., $\bar{\kappa}=\kappa_S=\kappa^{PM}$. However, it is also possible to dynamically tune the power mode depending on the state $S$, and this will yield diversified values of $\bar{\kappa}$.

Since the input rate is the inverse of the delay, we take
\begin{equation}
     \label{eq:q_lim}
     q_{\rm \textit{lim}} = \min \left\{q_{\rm \textit{lim}}^{\rm energy}, 1/\bar{\kappa}\right\}
\end{equation}
to also consider the processing delay. The value $q_{\rm \textit{lim}}$ represents the maximum job arrival rate a single edge device can tolerate with the given energy harvesting distribution. For any layer $\ell$, the scheduling probability of device \mbox{$i\in\ell$} is (line \ref{line:long_term_normalization})
\begin{equation}
    \label{eq:long_term_prob}
    r_i = \frac{q_{{\rm \textit{lim}}, i}}{\sum_{j\in\ell}q_{{\rm \textit{lim}}, j}}.
\end{equation}
This scheduling probability is used to select a device from the available ones within each group to execute an inference run.

\noindent\textbf{Adaptive scheduling.} For the adaptive policy, we propose a heuristic that adaptively tunes the input rate MDF. At each time step, the procedure starts by setting the input MDF to the static long-term solution (line \ref{line:long_term}). Then, the input rate of those devices in the critical power mode $PM_1$ is decreased by a quantity proportional to $\alpha/N_{\ell}$ (line~\ref{line:update_q}), where $N_\ell$ is the number of devices in layer $\ell$ and $\alpha \in [0, N_\ell]$ is a tunable parameter. Choosing the value $\alpha=|\{u \;|\; PM(u) = 1\}|$, i.e., the number of devices in the power mode $PM_1$, allows assigning to devices $v\;|\;PM(v) > 1$ a rate proportional to their cardinality in the considered layer. The obtained vector is finally re-normalized to obtain a valid MDF (line~\ref{line:normalization}).

\section{Experimental Results and Analysis}
\label{sec:results}

As a case study, we implement an LLM block with 100 encoder layers and 100 decoder layers, each employing 100 attention heads. We conduct inference on the AGX Orin device for an input with size \mbox{(64 x 16 x 512)}, measuring both energy consumption and processing time across various power modes. Accurate power readings are obtained using the method described in~\cite{calibrate}. We observe that for power modes of 15~W, 30~W, 50~W, and 60~W, the average processing times and energy consumption are approximately (300~s, 26~kJ), (200~s, 22~kJ), (205~s, 23.5~kJ), and (100~s, 23~kJ), respectively. The 50~W power mode exhibits about the same processing time as the 30~W mode, yet it surpasses the 30~W mode in terms of energy consumption. Therefore, due to its inefficiency in this task, we opt not to utilize it. Based on these empirical findings, in the collaborative inference simulation, a battery with a capacity of 100~kJ is chosen for each node and the time step is \mbox{$\delta = 100~s$}. Every node is associated with a uniform distribution representing energy arrivals, with a \emph{distinct mean}. The network topology consists of three groups of devices with three nodes each. For simplicity, each group hosts the same LLM block mentioned above. 
Communication times between devices are disregarded, and devices within each group are assumed to be fully connected to those in the next one. 

\begin{figure}
    \centering
    \subfloat[\label{fig:powermodes}]{\resizebox{.455\columnwidth}{!}{% This file was created with tikzplotlib v0.10.1.
\begin{tikzpicture}
\definecolor{seagreen6312776}{RGB}{63,160,76}
\definecolor{steelblue31119180}{RGB}{31,119,180}
\definecolor{mediumblue2742204}{RGB}{27,62,214}
\definecolor{forestgreen4416044}{RGB}{44,160,44}
\definecolor{darkorange25512714}{RGB}{255,127,14}
\begin{axis}[xtick pos=left,ytick pos=left,
    ybar,
    enlargelimits=0.15,
    legend style={nodes={scale=0.82}},
      xlabel={Power mode},
    symbolic x coords={15~W,30~W,60~W,dynamic},
    xtick=data,
    nodes near coords,
    nodes near coords align={vertical},
    ymajorgrids
    ]
\addplot[fill=darkorange25512714!50,draw=black!70] coordinates {(15~W,31) (30~W,45) (60~W,58) (dynamic,47)};
\addplot[fill=mediumblue2742204!43,draw=black!70] coordinates {(15~W,89) (30~W,42) (60~W,16) (dynamic,60)};
\legend{nr. of jobs completed in $T=100\delta$, average battery level (\%)}
\end{axis}
\end{tikzpicture}}}
    \hspace{0.1cm}
    \subfloat[\label{fig:xilim}]{\resizebox{.52\columnwidth}{!}{% This file was created with tikzplotlib v0.10.1.
\begin{tikzpicture}

\definecolor{crimson2143940}{RGB}{214,39,40}
\definecolor{darkgray176}{RGB}{176,176,176}
\definecolor{darkorange25512714}{RGB}{255,127,14}
\definecolor{forestgreen4416044}{RGB}{44,160,44}
\definecolor{gray}{RGB}{128,128,128}
\definecolor{lightgray204}{RGB}{204,204,204}
\definecolor{steelblue31119180}{RGB}{31,119,180}

\begin{axis}[
legend cell align={left},
legend pos=south east,
legend style={
  fill opacity=0.8,
  draw opacity=1,
  text opacity=1,
  draw=lightgray204,
  font=\small
},
tick align=outside,
tick pos=left,
x grid style={darkgray176},
xlabel={Job input rate $q$},
xmin=0.1, xmax=0.75,
xtick style={color=black},
y grid style={darkgray176},
ymode=log,
ylabel={Power saving mode risk $\xi$},
ymin=0.00000001, ymax=0.1,%0.0132,
ytick style={color=black},
xmajorgrids,
ymajorgrids
]
\addplot [ultra thick, steelblue31119180]
table {%
0.05 9.18102044127435e-11
0.1 7.71100050245676e-09
0.2 1.40995893709709e-06
0.25 9.25603194661637e-06
0.3 4.60247988559741e-05
0.35 0.000182062473688489
0.4 0.000581103729300383
0.45 0.00149442270045101
0.5 0.0031071510037405
0.6 0.00787521655244386
0.7 0.0125566335920268
0.9 0.0171484778615168
};
\addlegendentry{Dynamic}
\addplot [ultra thick, darkorange25512714, dashed]
table {%
0.05 2.62144352952528e-13
0.1 1.49455753868398e-11
0.2 1.66604364467328e-09
0.25 8.62774684241345e-09
0.3 3.35553739631491e-08
0.35 1.0421743971502e-07
0.4 2.67579357954038e-07
0.45 5.80990600745583e-07
0.5 1.08617925836663e-06
0.6 2.56583901939904e-06
0.7 3.95357212486041e-06
0.9 4.19893981054068e-06
};
\addlegendentry{15~W only}
\addplot [ultra thick, forestgreen4416044, dotted]
table {%
0.05 2.84053025644502e-12
0.1 3.08349828111962e-10
0.2 9.13849232490376e-08
0.25 7.73625688698415e-07
0.3 5.11386540143048e-06
0.35 2.83729586133303e-05
0.4 0.000137987885059182
0.45 0.00060427277228111
0.5 0.00241806471293364
0.6 0.0291928847561105
0.7 0.18795463512578
0.9 0.613015601215138
};
\addlegendentry{30~W only}
\addplot [ultra thick, crimson2143940, dash dot]
table {%
0.05 1.20639087671406e-08
0.1 9.266991214088e-07
0.2 0.000141390412431659
0.25 0.000881307163699844
0.3 0.00428330365635548
0.35 0.0169172578075349
0.4 0.0538637846124478
0.45 0.133511537587466
0.5 0.25416269711689
0.6 0.510544081203556
0.7 0.700983342567607
0.9 0.93039465869065
};
\addlegendentry{60~W only}
\addplot [very thick, black, dashed, forget plot]
table {%
0.05 0.01
0.1 0.01
0.2 0.01
0.25 0.01
0.3 0.01
0.35 0.01
0.4 0.01
0.45 0.01
0.5 0.01
0.6 0.01
0.7 0.01
0.9 0.01
};
\addplot [only marks,draw=gray,mark=*,mark options={fill=steelblue31119180,scale=2}]
table{%
x  y
0.646 0.01
};
\addplot [only marks,draw=gray,mark=square*,mark options={fill=darkorange25512714,scale=1.75}]
table{%
x  y
0.333 7e-08
};
\addplot [only marks,draw=gray,mark=triangle*,mark options={fill=forestgreen4416044,scale=2.5}]
table{%
x  y
0.50 0.0024
};
\addplot [only marks,draw=gray,mark=diamond*,mark options={fill=crimson2143940,scale=2.5}]
table{%
x  y
0.33 0.01
};

\end{axis}

\end{tikzpicture}}}
    \caption{\textbf{Power modes study.} %Analysis of the different power modes available on the Jetson Orin device.  
In (a), we observe the effects on job throughput and energy savings for both fixed and dynamic power modes. (b) illustrates the maximum job arrival rate while keeping the risk of entering the power saving mode under a user-defined threshold $\xi_{\rm \textit{lim}} = 0.01$.}
    \label{fig:power}
    %\vspace{-0.6cm}
\end{figure}
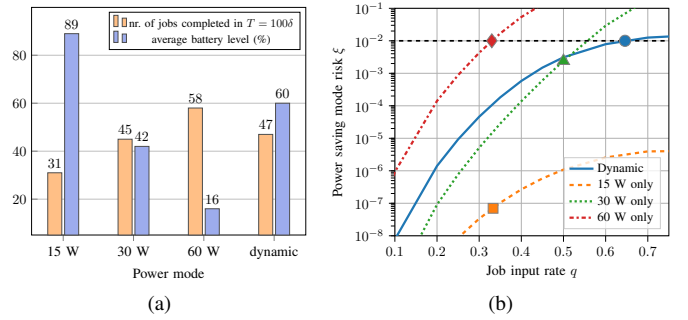

We evaluate the effectiveness of different power modes in facilitating decentralized LLM inference. Four power modes are compared: conventional 15~W, 30~W, and 60~W modes, along with a ``dynamic" power mode adapting to the device's current energy level. For the dynamic power mode, through manual exploration, we identified battery thresholds of 40\% and 60\% as optimal for transitioning from power mode 15~W to 30~W and from 30~W to 60~W, respectively. 
Fig.~\ref{fig:powermodes} compares the number of completed jobs and the average battery level in a simulation of length $100\delta$ for a single Orin device. The 15~W power mode, corresponding to $PM=1$, allows the device to save energy (89\% of battery) while having the lowest throughput, with only 31 completed jobs, as the mode is characterized by $\kappa=3$. Fixing the power mode to the highest possible, i.e., $PM=3, \kappa=1$, with 60~W, gives the dual result: we have the highest job throughput, with 58 completed jobs, but the battery is low (16\%) and the risk of entering the power saving mode is high. Using $PM=2, \kappa=2$ at 30~W is the best tradeoff considering fixed strategies, with 42\% of battery and 45 completed jobs. However, using the proposed dynamic power mode, with $PM\in\{1,2,3\}$ depending on battery level, job throughput increases (47 completed jobs) while having an 18\% average battery improvement concerning the 30~W fixed power mode.

The maximum input rate $q_{\rm \textit{lim}}$ is computed by 
applying Brent's method 
(see Section~\ref{sec:distributed_control}). Specifically, we calculate the risk $\xi_{\rm \textit{lim}}$ that the battery drops below the threshold $E_{\rm th}=10\%$, triggering the activation of the power saving mode.
Our objective is to find the highest input rate $q_{\rm \textit{lim}}$ to keep the risk less than the probability threshold $\xi_{\rm \textit{lim}}=0.01$. The results of this experiment are shown in Fig.~\ref{fig:xilim}. The lowest risk is attained by using 15~W, while the highest risk is by using the maximum power of 60~W. % 
For the 15~W and 30~W strategies, the threshold $\xi_{\rm \textit{lim}}$ (denoted by the horizontal dashed line) cannot be reached, as processing time constraints dominate in this case. In fact, the $q_{\rm \textit{lim}}$s (from Eq. \eqref{eq:q_lim}) are $1/3$ (the orange square) and $1/2$ (the green triangle), respectively. For the 60~W power mode, there are no time constraints, as it entirely processes a job in a single slot of duration $\delta$. The energy risk threshold is reached at $q_{\rm \textit{lim}}\approx 0.33$ (red diamond). For the dynamic power mode, the energy threshold is hit at $q_{\rm \textit{lim}}\approx 0.64$ (blue circle). This is also well approximated by $1/\bar{\kappa}$, i.e., the average maximum rate that can be tolerated for a stable input queue. According to these results, using a dynamic power mode allows enduring a higher input rate, while controlling the downtime risk below $\xi_{\rm \textit{lim}}$.
In other words, the dynamic power mode effectively manages energy resources while optimizing the throughput for decentralized inference over edge networks. %

\begin{figure}
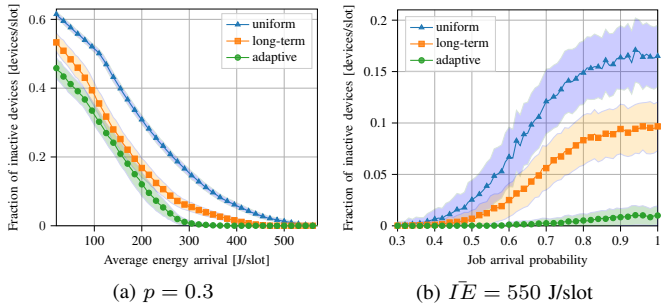

    \centering
    \subfloat[$p=0.3$\label{fig:down-energy}]{\resizebox{.478\columnwidth}{!}{\input{Downtime-Energy}}}
    \hspace{0.1cm}
    \subfloat[$\bar{IE}=550$ J/slot\label{fig:down-job}]{\resizebox{.497\columnwidth}{!}{\input{Downtime-Ja}}}
    \caption{\textbf{Fraction of inactive devices.} Share of devices in power saving mode due to a lack of harvested energy {\it vs.} average energy arrivals (a) and job arrival probability (b).}%\vspace{-0.3cm}
    \label{fig:downtime}
\end{figure}

For subsequent experiments, we use the dynamic mode as the power strategy and investigate the scheduling policies outlined in Section~\ref{sec:distributed_control} through simulated inference runs of the system. 
We repeat system runs for 1,000 iterations per experiment and depict the mean and standard deviation in the forthcoming plots. %
We explore how the fraction of devices in power saving mode changes concerning the average energy arrival (Fig.~\ref{fig:down-energy}) and the job arrival probability (Fig.~\ref{fig:down-job}). As can be seen, blindly applying a uniform distribution among available devices for the job assignment ends up in having a sub-optimal solution in terms of downtime probability. This is because the uniform distribution does not consider the differences in energy arrival for the scheduled devices.

In contrast, long-term selects the scheduling probabilities set by considering the average optimal solution according to the semi-Markov model developed, so that the scheduling distribution is unbalanced towards the richest devices for energy income. This allows long-term to outperform the naif uniform distribution: the downtime risk is reduced to 15\% when varying the energy arrivals and is consistently halved when varying the job arrival rate. 
Using the adaptive method, a reduction of up to 10\% in the downtime probability is observed compared to long-term when varying both energy and job arrival rates. Notably, with the chosen energy settings, even when one job arrives at each time slot (i.e., the processing is always active), adaptive can keep the downtime probability as low as about 1\%.

\begin{figure}
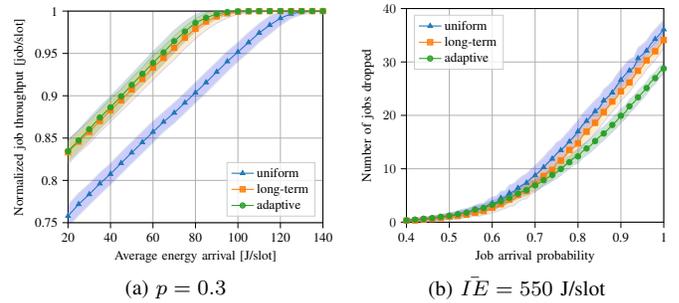

    \centering
    \subfloat[$p=0.3$\label{fig:throughput}]{\resizebox{.498\columnwidth}{!}{\input{Throughput-Energy}}}
    \hspace{0.1cm}
    \subfloat[$\bar{IE}=550$ J/slot\label{fig:dropped}]{\resizebox{.478\columnwidth}{!}{\input{Dropped-Ja}}}
    \caption{\textbf{Edge network processing capacity.} The processing capacity of the proposed decentralized inference system is evaluated as normalized job throughput or fraction of dropped jobs {\it vs.} energy arrivals (a) and job arrivals (b).}%\vspace{-0.3cm}
    \label{fig:jobs}
\end{figure}

In Fig.~\ref{fig:throughput} we show the normalized job throughput (i.e., the number of jobs processed by the system concerning the total input jobs) with respect to the average energy arrivals. The scheduling methods based on our semi-Markov model outperform the uniform scheduling: They increase the job throughput by about 10\% when the energy arrival is relatively low. 
Compared to long-term, using the adaptive scheduling policy allows the system to further gain some processing capacity on average (about 2\%). Fig.~\ref{fig:dropped}, instead, shows the dual metric, i.e., the number of jobs dropped by the system, when varying the job arrival probability. The three strategies provide similar results up to the input rate $p\approx 0.65$, where they enter a linear behavior with a similar slope. However, using uniform scheduling, we enter this region faster than the proposed methods, which have slower elbows (transition regions). On average, long-term and adaptive drop about 3 and 7 jobs less than uniform, respectively.

\section{Concluding Remarks}
\label{sec:remarks}

This work presented a semi-Markov model for edge computers equipped with energy harvesting. The model was tested with a decentralized LLM inference task, considering time and energy constraints, and using simple scheduling algorithms. The experiments were performed using parameters from real energy and time measurements of a Jetson Orin. 
Using the proposed semi-Markov model and the derived scheduling policies, the processing performance has been improved while effectively reducing the shutdown risk. The method presented in this paper can be adapted to similar scenarios with different parameter settings, allowing its application across various contexts and conditions.

\bibliographystyle{IEEEtran}
\bibliography{Ref}

\end{document}